# Elastomeric 2D Grating and Hemispherical Optofluidic Chamber for Multifunctional Fluidic Sensing

Zhida Xu,[1,2*] Xinhao Wang,[1,2] Kevin Han,[2] Shuo Li,[1,2] and G. Logan Liu[1,2]

[1]*Micro and Nanotechnology Laboratory, University of Illinois at Urbana-Champaign, Urbana, IL 61801, USA*
[2]*Department of Electrical and Computer Engineering, University of Illinois at Urbana-Champaign, Urbana, IL 61801, USA*
*\*zhidaxu1@illinois.edu*

We present an optofluidic sensor based on an elastomeric two-dimensional (2D) grating integrated inside a hemispherical fluid chamber. Laser beam is diffracted before (reflection) and after (transmission) going through the grating and liquid in the dome chamber. The sensing mechanism is investigated and simulated with a finite difference time domain (FDTD) based electromagnetic (EM) method. For experiment, by analyzing the size, power and shape of the 2D diffraction patterns, we can retrieve multiple parameters of the liquid including the refractive index, pressure and opacity with high sensitivity. We demonstrate that glucose concentration can be monitored when mixed in different concentrated phosphate buffered saline (PBS) solution. The free-solution binding of bovine serum albumin (BSA) and anti-BSA IgG is detected with this optical sensor. This low-cost, multifunctional and reliable optofluidic sensor has the potential to be used as monitor of biofluid such as blood in hemodialysis.

## 1. INTRODUCTION

Monitoring of biofluid is of dominant importance in bioassays such as diagnosis and treatment of diseases. Take blood monitoring in hemodialysis for example: different parameters of blood need to be extracted, including arterial and venous pressure, concentration of dialysate and waste.[1] As a result, multiple monitors are needed thus making the system bulky and expensive. Due to high sensitivity and precision, optical sensors are incorporated into fluid channel for real-time measurement of the fluid's properties. For example, laser interferometer can be used to measure refractive index as well as concentration of glucose solution. [2] Step-index optical fiber can be used as refractive index sensor of fluids.[3] Back-scattering interferometry can be used to monitor free-solution and label-free molecular interaction. [4] Whole blood in hemodialysis can be monitored with transmission and diffuse reflection spectroscopy.[5] In additional to the opacity and refractive index, the mechanical properties of fluid such as pressure can be measured by deformable optical elements like polymer diaphragms.[6, 7] Deformable or elastomeric gratings are used as photothermal detectors[8] and micro-strain sensors[9], refractive-index sensors for high-refractive-index solid[10], monitor of local pressure in microfluidic devices[11] and fluidic actuation[12]. The advantages of using deformable grating as optofluidic sensor include high sensitivity, high spatial resolution, high reliability, low cost and simplicity in manufacturing. [13]

In this article, we demonstrate an optofluidic sensor based on a 2D elastomeric grating integrated in a hemispherical fluid chamber (EGS sensor). It is a prototype of multi-function fluid sensor to monitor the refractive-index, opacity and pressure of fluid in real time. The sensor is completely made of polydimethylsiloxane (PDMS). PDMS elastomer can be easily molded into surface relief grating and it is transparent in the

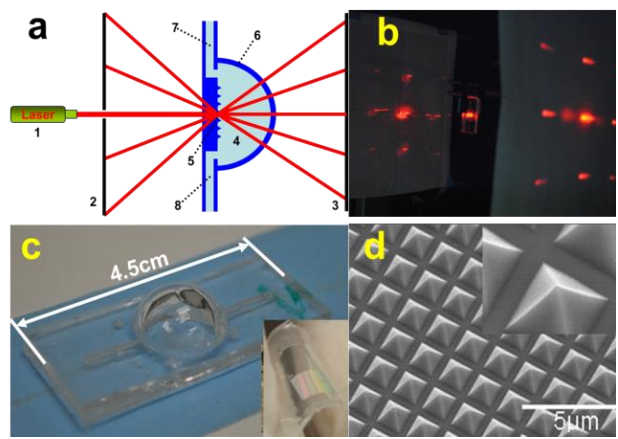

Fig. 1. (a) Sketch of the diffraction sensing system. ① Laser, 633nm, 4mW ② Backward diffraction screen. ③ Forward diffraction screen. ④ Liquid. ⑤ Elastomeric 2D grating. ⑥ Half-dome chamber. ⑦ Fluid inlet. ⑧ Fluid outlet. (b) Photograph of the diffraction sensing system in measurement. (c) Photograph of the EGS sensor. The inset shows deformed 2D grating. (d) SEM image of micropyramids array structure on the 2D grating.

ultraviolet–visible range (300 – 800 nm). [14] Fig. 1(a) is a sketch of the overall setup. The 2D grating in the middle of the half-dome chamber is the diffraction sensing element. It is a 2D array of pyramids made by nanoreplication of 2D inverted nanopyramids array on silicon mold with PDMS. Previously we have demonstrated SERS substrate can be produced in this rapid and low cost method.[15] The beam from a semiconductor laser with the power of 4 mW and wavelength of 633 nm is diffracted, both forwardly (transmission) and backwardly (reflection).

## 2. MONITORING OF REFRACTIVE INDEX OF FLUID

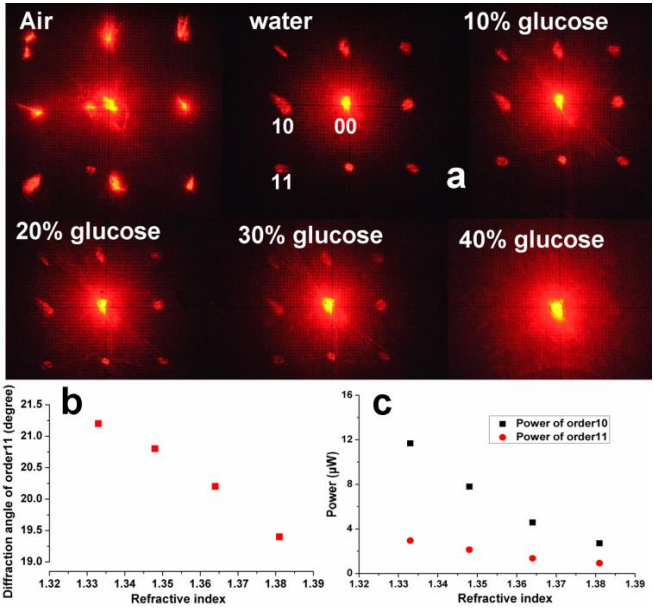

Fig. 2. (a) Forward diffraction patterns for EGS device containing air, water and glucose aqueous solution with concentration of 10%, 20%, 30% and 40% by mass. (b) Diffraction angle of order11 for EGS device containing glucose solution with concentration of 0% (water, n=1.33),10%(n=1.348), 20% (n= 1.364), 30% (n=1.381). (c) Average power of diffraction order 10 and 11 for glucose solution with concentration of 0% (water, n=1.33),10%(n=1.348), 20% (n=1.364), 30% (n=1.381).

The sensing principle of EGS sensor is based on the diffraction pattern of the grating. After the laser beam passes through the hole in the center of screen for display of backward diffraction pattern, it hits the 2D grating and gets diffracted both forwardly and backwardly. The diffraction equation for 1D grating is:

$$nd \sin(\theta_m) = m\lambda \qquad (1)$$

From equation 1, we can see that for fixed wavelength λ, grating constant d and diffraction order m, the diffraction angle $\theta_m$ depends on the refractive index n of medium on the diffraction side. For backward diffraction, the medium is air with n = 1. For forward diffraction, the medium is liquid in the hemisphere chamber thereby its refractive index n can be retrieved from the forward diffraction angle. The purpose of the hemisphere shaped fluidic chamber is to keep the angle of forward diffraction by ensuring normal incidence on the interface; otherwise the beam will be refracted to a different angle. As the 2D grating is elastomeric, it will be deformed under pressure. The inset in Fig. 1(c) shows the grating being bended by fingers. Once the grating is deformed, the backward diffraction pattern will change. The fluid pressure change can be monitored by the change in backward diffraction pattern. Fig. 1(c) is a photograph of the device with flexible 2D grating and hemisphere fluid chamber. We call it EGS device here. It is made by replication of inverse pyramids array with PDMS, for which the process was demonstrated before.[15] We made the thickness of grating film as thin as 0.1mm so it is quite sensitive to pressure change. Only a little curing agent (1:20) is used to make it soft. The hemisphere chamber and microfluidic channels are also made with PDMS by molding. More curing agent (1:5) is used to make them stiffer than the grating film. After all they are glued together and sealed also with PDMS.

To demonstrate measurement of refractive index, we prepared glucose solutions in different concentrations. The refractive indices are looked up in [16] and tabulated in Table 1. Two parameters including diffraction angle and diffraction power can be used to measure refractive index. All the experiments are done at room temperature of 25 °C. Fig.2 (a) shows the forward diffraction patterns when EGS sensor is filled with air, water and glucose solutions in different concentrations. Fig.2(b) shows diffraction angles calculated from Fig. 2(a). Fig.2(c) is the plot for averaged power of order 10 and 11, measured with a silicon photodetector. In this setup, the EGS is 16 cm and 19.5 cm from the forward diffraction screen and backward diffraction screen respectively. As the concentration increases, the diffraction pattern become smaller in size and the power of high-order diffraction is weaker. At the mass concentration of 40%, the diffraction pattern disappears because the refractive index of solution is quite close to that of PDMS (n = 1.41). Table 1 lists measured parameters for glucose concentrations from 10% to 20%. From Fig. 2 and Table 1, we can see using power to monitor refractive index is more sensitive compared with using diffraction angle. By taking percentage change of target parameter (from n=1.333 to n=1.381) to characterize the relative sensitivity, we found that the sensitivity of using power is 2075%/RIU while that of using angle 193.3%/RIU.

## 3. FDTD SIMULATION OF DIFFRACTION

To calculate the change of forward diffraction pattern and power with the change of refractive index, we use finite difference time domain (FDTD) method to numerically solve Maxwell equations of electromagnetics for simulation. The commercial software Lumerical FDTD solution is used for FDTD simulation. We built a 3D model with a large 2D array (17×17) of pyramids (Fig 3.(a)). The size and period of the pyramids array are set according to measurement by SEM in Fig.1(d). According to dimensions measured with SEM, the pitch of the pyramids array is 2 µm, and the base lateral length of pyramids is 1.58 µm.[15, 17] The angle between the flat plane (100) and inclined surface (111) is 54.7° due to KOH anisotropic etching so the height of the pyramid is 1.15 µm[18]. We first tried a smaller array of pyramids with periodic boundary condition and Bloch boundary condition but the result of diffraction did not turn out good. Then we found that a larger array of pyramids with perfect matched layer (PML) boundary condition could result in more clear pattern of diffraction which matched with the experimental result. Two plane wave sources with orthogonal polarizations are placed beneath the pyramids array (Fig.3(b)). Both are 633 nm in wavelength and have the same phase. The reason we use two orthogonally polarized wave sources is to make the diffraction pattern symmetric. The laser we used in experiment is unpolarized thus the diffraction pattern is symmetric. We set the refractive index of the pyramids array to be 1.41 as PDMS and change the refractive index of environment on the pyramids' side from 1 to 1.4 with 24 points. Every time the refractive index of environment is changed, we run the simulation to solve the diffraction pattern and power in the farfield.

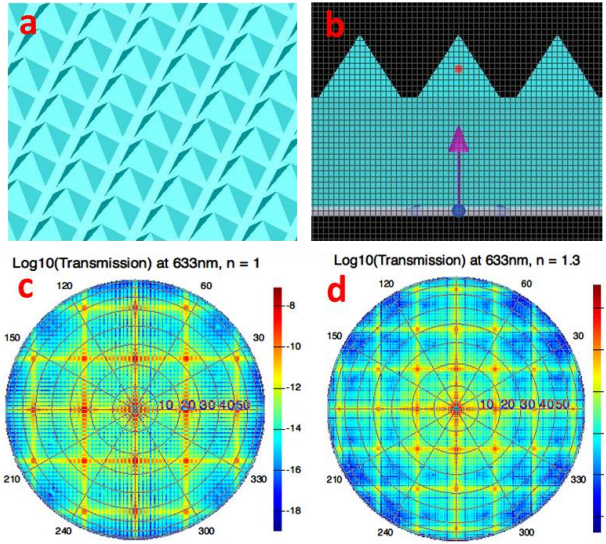

Fig. 3. Model of FDTD simulation for forward diffraction of pyramids 2D array. (a) Perspective view of the pyramids 2D array. (b) Cross-section view of pyramids 2D array and two orthogonal plane wave sources. The pink arrow indicates the direction of propagation while the blue arrows indicate the polarizations. (c) Calculated forward diffraction pattern in the far field when the refractive index of environment n is 1. The (d) Calculated forward diffraction pattern in the far field when n = 1.3. The relative intensity of electric field as $|E|^2$ is plotted in Log10 scale with color indicated by the color bar.

Fig. 3(c) and (d) show the forward diffraction patterns of pyramids 2D array in the farfield when the refractive index of the environment is 1 (c) and 1.3 (d) respectively. The intensity of electric field $|E|^2$ is plotted in color. Red represents high intensity while blue represents low intensity. Comparing Fig. 3(c) with Fig. 3(d), we can see as the refractive index n of environment increases from 1 to 1.3, the diffraction angle becomes smaller and the power of non-zero order diffraction becomes weaker. This is consistent with the experiment results in Fig.2. The data retrieved from FDTD simulation results is plotted in Fig. 4.

Fig.4 (a) shows the simulated diffraction angle of order 10 and order 11 with n from 1 to 1.4. The simulated results in Fig. 4(a) match well with experiment results in Fig. 2(b) and Table 1. The diffraction angle of order 11 drops from 28.5 ° to 19.3° as n increases from 1 to 1.4. The diffraction angle of order 10 is always proportional (1/√2 in tangent value) to that of order 11 due to geometrical relation. The simulated power of diffraction order 10 and 11 is plotted in Fig. 4(b). We take the intensity of electric field $|E|^2$ to represent the power. The amplitude of incident electric field is 1 thus the value of $|E|$ we take is a relative number to incident field. Interesting thing we observed in Fig. 4 (b) is that both order 10 and 11 show a maxima around n = 1.1. The power first increases then decreases. We come up with an explanation for this trend of simulation. That is, for the diffraction power, there is a competition between transmission and diffraction. When n is low, it is far from the refractive index of PDMS (1.41). In this case the reflection at the interface of the environment and PDMS grating is high due to an abrupt change of the refractive index at the interface. As reflection is high, the transmission is low. So as n increases closer to that of PDMS, the reflection drops and transmission increases. This gives the increasing trend of the diffraction power before it reaches the maxima. But as n increases closer to the refractive index of PDMS, the grating effect becomes weaker thus decreases the diffraction power. As a result, when n is low, the effect of high reflection is dominant so the diffraction power increases with n. When n is high, the weakening of grating effect is dominant so the diffraction power decreases with n. At the maxima is where reflection and diffraction switch their dominance. Fig. 4 (c) shows a monotonically increasing curve of power of order 00 as n increases. In this case, for the order 00, the non-diffracted order, both reflection and diffraction contribute to its increasing trend. As n increases closer to 1.41, the transmission increases and diffraction decreases so more light are transmitted through this non-diffracted order 00. The trends of diffraction order 10 and 11 with n are consistent with the experiment results shown in Fig. 2 and Table 1. From n=1.3 to n=1.4, which is the refractive index range of glucose solution, both experiment and simulation show the power drop of diffraction order 10 and 11 by several times, demonstrating the high sensitivity of ECS sensor for refractive index.

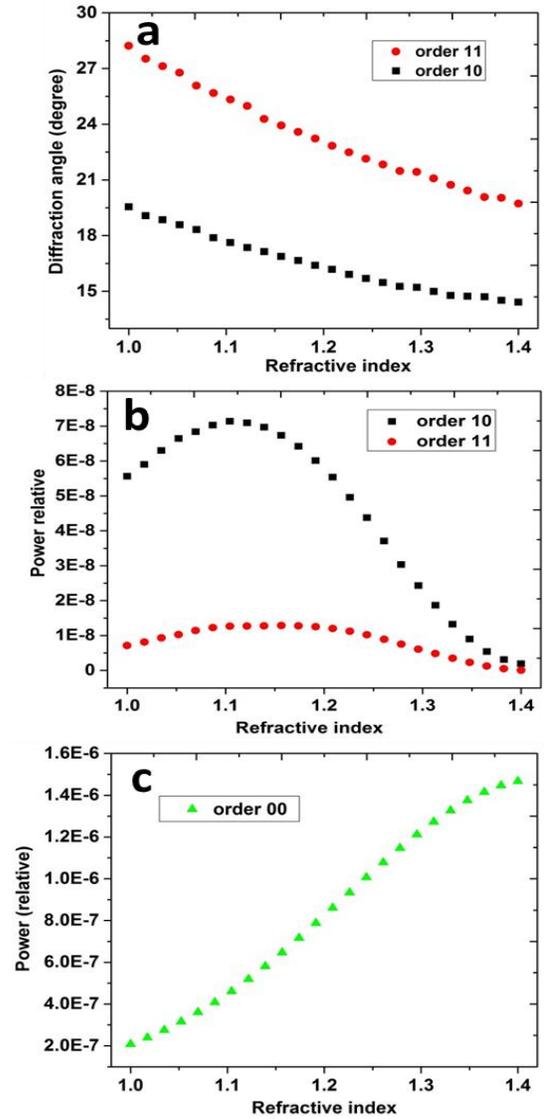

Fig. 4. FDTD simulation results of diffraction pattern of pyramids 2D array with different refractive index of environment from n=1 to n=1.4. (a) Diffraction angles of order 10 and 11 from n=1 to n=1.4. (b) Diffraction power of order 10 and 11. (c) Diffraction power of order 00 from n =1 to n = 1.4.

Table 1. Refractive index, power of 3 orders, diffraction angle of order 11 for glucose solution with different concentrations.

| Glucose C | n(RI) | Power of order $\bar{0}\bar{0}$ | Power of order $\bar{1}1$ | Power of order $\bar{1}\bar{0}$ | Angle of order $\bar{1}1$ |
|---|---|---|---|---|---|
| Air | 1 | 0.9mW | 33.78µW | 168.3µW | 28.3° |
| 0% | 1.333 | 2.582mW | 2.95µW | 11.68µW | 21.2° |
| 10% | 1.348 | 2.81mW | 2.13µW | 7.8µW | 20.8° |
| 20% | 1.364 | 2.78mW | 1.36µW | 4.57µW | 20.2° |
| 30% | 1.381 | 2.67mW | 0.93µW | 2.71µW | 19.4° |
| 40% | 1.4 | 2.5mW | \ | \ | \ |
| 50% | 1.42 | 2.5mW | \ | \ | \ |

## 4. MONITORING GLUCOSE CONCENTRATION IN PBS SOLUTION

In order to investigate whether glucose concentration can be monitored with EGS sensor when other molecules are mixed in the solution, we put glucose in phosphate buffered saline (PBS) solution, which is a water-based buffer solution containing sodium chloride and sodium phosphate commonly used in biological research. The PBS solution we used is BioWhittaker 10X (0.1M) PBS purchased from Lonza Group Ltd. We diluted the 10X PBS solution into 1X (0.01M) and 0.1X (0.001M) PBS solution. Then we prepared 10%, 20%, 30% and 40% glucose solution by weight with 0.1X, 1X and 10X PBS solution as solvent respectively. From Fig. 2 and Table 1 we have learnt that the most sensitive features changing with refractive index is the power of diffraction order $\bar{1}\bar{0}$ and 11 while the diffraction angle and order $\bar{0}\bar{0}$ are not that sensitive. For experiments with glucose in PBS solution, we only looked at power of diffraction order $\bar{1}\bar{0}$ and 11. Fig.5 shows the power of diffraction order $\bar{1}\bar{0}$(Fig. 5(a)) and 11(Fig. 5(b)) with different glucose and PBS concentration. At 40% glucose, the diffraction pattern is hardly visible but we still measured the power at the place where the diffracted beam supposed to be. That is because the diffraction exists even though it is hardly to see by eye. To correlate the measured power with the refractive index of the liquid, we use the data in Table 1 to fit a curve of refractive index to diffracted power. By looking at the trend in Fig. 2, we use second degree polynomial for curve fitting. We got fitting polynomial $0.00622 x^2 - 0.04712 x + 1.41859$ for order 11 and polynomial $0.00048517 x^2 - 0.01237 x + 1.41235$ for order $\bar{1}0$. Based on these fitting polynomials, the refractive index corresponding to diffracted power is calculated and plotted on the right axis of Fig. 5. The trend of experiment result is clear for both order $\bar{1}0$ (Fig. 5(a)) and order 11 (Fig. 5(a)). For the same concentrated PBS solution, the diffraction power decrease dramatically with the increase of glucose concentration. For the same concentration of glucose, the diffraction power also decreases with the increase of PBS buffer concentration. The results proved that the glucose concentration is still detectable if other molecules are mixed in the solution.

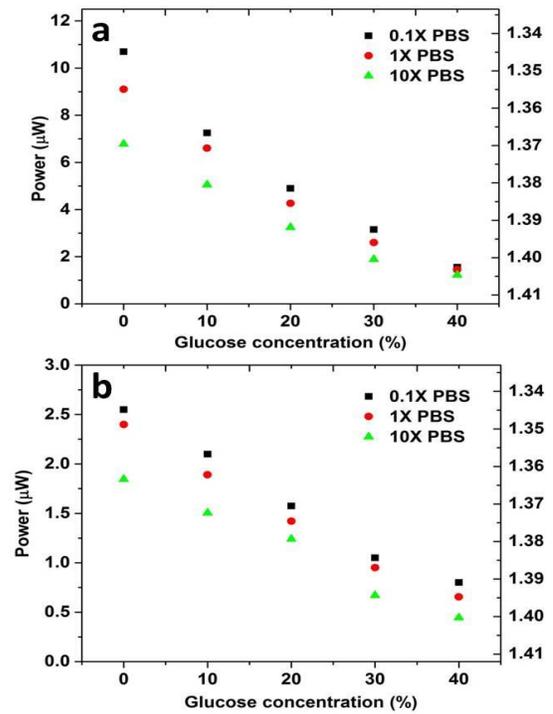

Fig. 5. Power and corresponding refractive index of diffraction order 10(a) and order 11(b) for glucose (0%,10%,20%,30%,40%) in PBS solution(0.1X, 1X, 10X) with different concentrations.

## 5. MONITORING PRESSURE CHANGE AND FLUID ACTUATION

We use backward diffraction pattern to monitor the pressure of the fluid. The backward diffraction pattern is almost not affected by the refractive index of liquid in the chamber as we confirmed by experiment. The film with the grating is deformed when pressure changes, so the diffraction pattern will shift. Fig. 6 is a schematic to show how pressure is monitored. We use a commercial pressure sensor connected in series with EGS sensor so

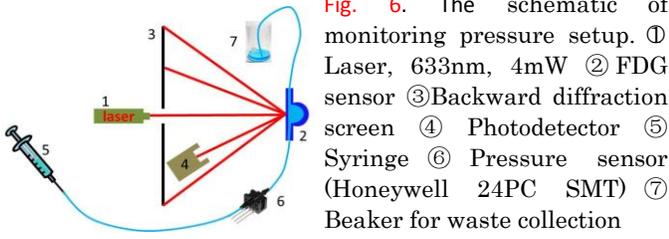

Fig. 6. The schematic of monitoring pressure setup. ① Laser, 633nm, 4mW ② FDG sensor ③Backward diffraction screen ④ Photodetector ⑤ Syringe ⑥ Pressure sensor (Honeywell 24PC SMT) ⑦ Beaker for waste collection

they undergo the same pressure. Then a syringe is used to control the pressure.

The syringe is pushed or pulled to create pressure change. The commercial pressure sensor measures the pressure change by voltage sensitivity of 16.7 mV/psi and we compare it with EGS sensor. For EGS sensor to measure pressure, we put a photodiode in front of the backward diffraction screen to measure the power of diffraction order $\bar{1}0$ beam. The power is measured by Thorlabs PM130D powermeter controlled with LabVIEW program with a sampling rate of 5Hz. When pressure changes, the grating film is deformed thus the diffraction beam is shifted. In this way, small pressure change can be detected very sensitively. Fig. 7 compares the pressure sensing results of commercial sensor and EGS sensor in 90 seconds. The drops around 15s, 25s, 52s and 76s are large pressure changes (>1.5 psi) while the fluctuation from 55s to 75s is a small pressure fluctuation (<0.5 psi) we created on purpose. We can see the EGS sensor reading makes more drastic change under pressure change, especially at small pressure fluctuation, with a sensitivity of about 5 μW/psi. But at large pressure change, the reading of EGS sensor drops rapidly to zero so it falls out of range. The EGS sensor is proved to be better for qualitative detection of small pressure change than quantitative measurement.

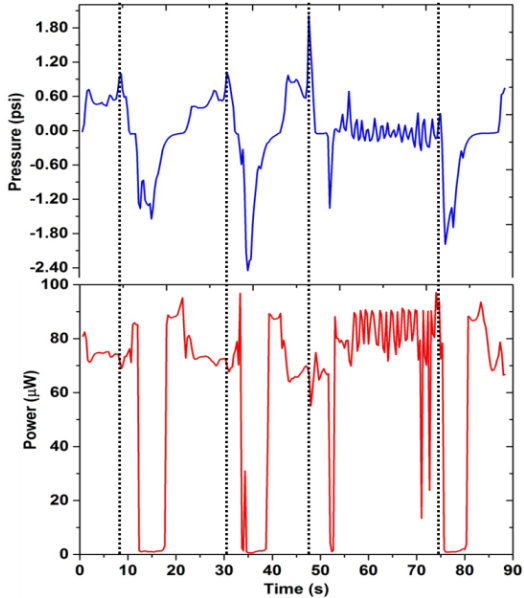

Fig. 7. Fluid pressure monitoring plot in 90 seconds for commercial fluid sensors (blue curve) and EGS sensor (red curve). Dashed lines indicate when major pressure change happened.

## 6. PRINCIPLE OF GRATING DEFORMATION BY FLUID ACTUATION

Now let us do a simple analysis of how the backward diffraction pattern will be affected by the deformation of grating when we apply and release pressure of the fluid chamber. The elastomeric grating membrane is deformed due to the change of volume of liquid in the fluid chamber therefore the spacing between pyramids is changed. Fig. 8(a) on next page is a cross-sectional sketch of grating membrane without deformation. The initial length of this segment of grating is $s_0$ and initial spacing, or grating constant between adjacent pyramids is $d_0$. Injecting additional fluid deforms and elongates the grating. Fig. 8(b) is a cross-sectional sketch of the grating membrane after being deformed by fluid injection when pressure is applied. Because the fluid is on the side of the pyramids, the injection of fluid causes the grating to bend downwards. The bending curvature is exaggerated here for visual reason. Assuming the fluid is incompressible and others parts except the grating membrane are non-deformable, the injected volume of fluid $\Delta V$ is equal to the volume by distended membrane, shown as the shaded area in Fig. 8(b). Because $\Delta V$ is small compared to the volume of the whole chamber, let us approximate the distended grating membrane as a spherical surface. [19] Incident beam is normal to the center of grating membrane. The deformed grating length s can be expressed as: [12, 20]

$$s = R\phi = R\cos^{-1}(1 - \frac{2r_o^2}{R^2}) \qquad (2)$$

All the parameters in equation (2) are indicated in Fig. 8. The angle $\Phi$ is written in terms of radius of curvature R and aperture radius $r_o$. The bending curvature is exaggerated on Fig. 8 for visual effect but actually the curvature is very small so that we assume the incident light remains normal to the grating plane. The variation of R as a function of $\Delta V$ is obtained by rearranging a relationship between $\Delta V$ and R for spherical membranes: [12, 20, 21]

$$\frac{12\Delta V}{\pi}R^3 - 3r_o^4 R^2 - \left(r_o^6 + \frac{9\Delta V^2}{\pi^2}\right) = 0 \qquad (3)$$

Equation (3) can be numerically solved for R to determine segment length of grating s at given $\Delta V$. In general, for fixed apertures, as the elastic surface expands, the radius of curvature is reduced.[19] According to the grating equation (1), in each diffraction direction, the diffraction angle $\theta_m$ of m-th order diffraction is:

$$m\lambda = d_s \sin(\theta_m) \qquad (4)$$

In which $d_s$ is effective spacing between adjacent pyramids, or grating constant. Fluid actuation modifies $d_s$ as a function of the arc length: [12, 20, 21]

$$d_s = \frac{s}{s_o}d_o \qquad (5)$$

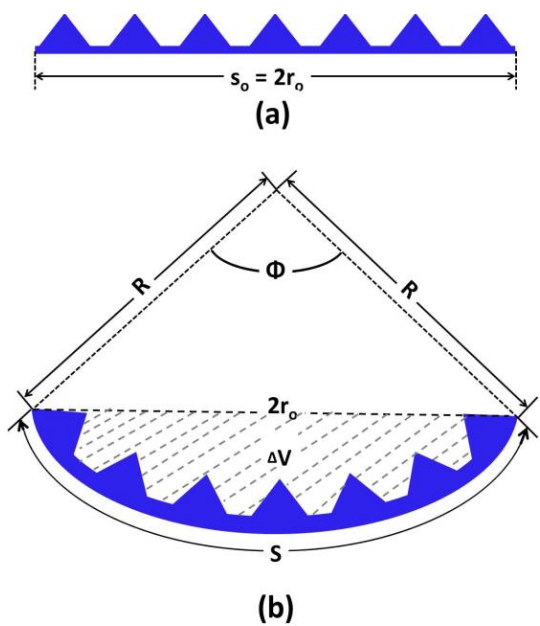

Fig. 8. Deformation of elastomeric pyramids 2D grating by injection of fluid. (a) Cross-section of initial grating membrane before injection of fluid. (b) Cross-section of deformed grating member after fluid of volume ΔV being injected. Assuming uniform strain along the grating membrane, the modified spacing between pyramids can be calculated from the effective radius of curvature R.

under the assumption that strain is uniform across the grating membrane. With the grating constant changed by fluid actuation, the diffraction angle and pattern will also be changed.

So far we have demonstrated the capability of EGS sensor for sensing of refractive index and monitoring of pressure change and fluidic actuation. To show an example of using EGS sensor for monitoring bio-fluid, we performed free-solution protein-antigen and antibody binding. Free-solution means the binding reaction happens within the fluid rather than on the surface, so no surface functionalization is needed. We prepared both bovine serum albumin (BSA) and anti-BSA immunoglobulin G (IgG) with concentration of 10 μM for both. Both antigen and antibody are purchased from Sigma-Aldrich. We use two syringes, one for BSA and the other for IgG, and connect two syringes with a Y-connector. So the binding happens before the fluid enters the half-dome chamber of the EGS sensor. We suppose the antigen-antibody binding will change the fluid's refractive index and opacity a little so forward diffraction is used to monitor the binding. We use a photodetector to measure the power change of the $\overline{1}0$ diffraction beam. The experiment was performed at the temperature of 37 °C (optimal binding temperature). Fig. 9 shows the power change before and after binding. The binding happened at around 60s and we can see a rapid drop in the power from 12.7 μW to 12 μW of diffraction order $\overline{1}0$, followed by power stabilization, which indicates an increase and then a new steady state value in refractive index of the fluid.

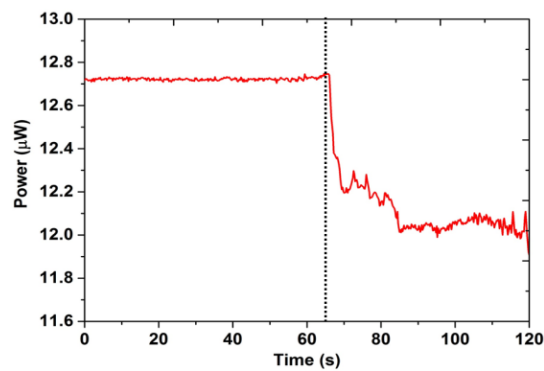

Fig. 9. Monitoring BSA and anti-BSA IgG binding with EGS sensor by power. The dashed line indicates when the BSA and anti-BSA solutions were mixed.

## 7. DISCUSSION AND PERSPECTIVE

One tradeoff for high sensitivity is the vulnerability to the ambient disturbance. During the experiments we found that small disturbance will adversely affect the measurement results, especially for the pressure sensing. For instance, if we touch the membrane with a tweezer while doing measurement, the results will change drastically. However, that is the tradeoff for high sensitivity. Therefore, before we did the measurement, we tightly fixed the EGS sensor on a stage with tape, and the whole setup is mounted on optical table to minimize the vibration. We also used extreme caution when doing the measurement so that other factors do not interfere with the results. When we re-design the sensor in the future, the most important thing to consider is how to make it immune to ambient disturbance while keeping its sensitivity. Using more rigid material to make other parts except the grating membrane should help.

Temperature is a very import ambient factor which may affect the diffraction of PDMS grating. Actually people do use elastomeric grating as thermal sensor.[8, 22, 23] The temperature will affect not only the geometry, density and refractive index of the pyramidal 2D grating but also the density, pressure and refractive index of the liquid in the chamber. It is a very complicated mechanism based on the effect of temperature change. So for all the refractive index and pressure characterization, we keep the room temperature at 25°C using the air conditioner in our lab. For the BSA-antiBSA binding experiment, we changed the room temperature to 37°C, which is the optimal temperature for BSA-antiBSA binding and kept all the setup and solutions at this temperature before we did the experiments. We observed little change of the diffraction pattern when temperature increased from 25°C to 37°C. However, changing the room temperature may not be the best way to accurately control the temperature of the device itself. It is possible for this EGS sensor to be used as highly sensitive temperature sensor as well. Now we are building a different setup based on a copper hotplate specifically to characterize the effect of temperature. But the work and analysis of temperature characterization will be quite lengthy and comprehensive and we will report in a future manuscript.

For the part of pressure sensing and fluid actuation, we analyzed the principle of how the grating membrane is deformed by fluid actuation in section 6. The change of grating constant as a function of injected fluid volume ΔV is derived under the assumption that other parts of the sensor except the grating membrane are strictly rigid. However, this model is somewhat over-simplified. Firstly, as the grating membrane is stretched, it will be thinner and the pyramids will widen and lower thus it will have different optical properties. Secondly, if the laser beam is not aligned right in the center of the grating membrane, the deformation of grating will cause the entire diffraction pattern to shift to one direction. But this shifting is actually more sensitive than the shrinking of the diffraction pattern thus by placing the beam slightly off the center we can make a more sensitive pressure monitor. In section 6, the deformation of grating is given as a function of injected volume ΔV. It is possible but complicated to derive the relation between deformation and pressure being applied. And the assumption that other parts of the chamber are rigid is not that accurate because they are also made of PDMS but thicker. We will do some simulation to analyze the deformation of the whole chamber under pressure being applied from the fluid inlet.

This BSA and anti-BSA IgG binding experiment is just a proof of concept of biological experiments being carried on with EGS sensor. To prove the capability of this sensor to be used in specialized biofluid experiments, we will test this sensor with more advanced experiments such as DNA hybridization[24] and enzyme-linked immunosorbent assay (ELISA)[25]. Our ultimate goal is to apply this EGS sensor in clinical use.

## 8. CONCLUSION

To summarize, we demonstrated a prototype of multi-function optical sensor based on diffraction pattern. It contains an elastomeric 2D grating and semi-spherical fluid chamber. Its sensing mechanism was investigated by FDTD-based electromagnetic simulation. It can be used to monitor the change of refractive index and pressure of fluid with high sensitivity. The concentration of glucose when mixed in PBS solution can be dynamically monitored. Detection of protein binding reaction in free solution was presented.